\documentclass[journal]{article}                                  

\usepackage{arxiv}
\usepackage{hyperref}
\usepackage{graphicx}          
\usepackage{amsmath} 
\usepackage{amssymb}  

\usepackage{amsthm}
\usepackage{color,soul}
\usepackage{flushend}
\usepackage{cite}

\usepackage{mathtools}
\usepackage{xcolor}
\usepackage{array,multirow}
\usepackage{algorithm} 
\usepackage{algorithmic}  
\usepackage[linesnumbered,algo2e,ruled,vlined,norelsize]{algorithm2e} 

\allowdisplaybreaks
\usepackage{tikz}
\usepackage{textcomp}
\usepackage{hyperref}
\usepackage{lipsum}

\newcommand\copyrighttext{%
  \footnotesize \textcopyright 2025 IEEE. This work has been accepted to the 2025 American Control Conference (ACC). Personal use of this material is permitted.
  Permission from IEEE must be obtained for all other uses, in any current or future 
  media, including reprinting/republishing this material for advertising or promotional 
  purposes, creating new collective works, for resale or redistribution to servers or 
  lists, or reuse of any copyrighted component of this work in other works. 
  DOI: \href{<http://tex.stackexchange.com>}{To be generated.}}
\newcommand\copyrightnotice{%
\begin{tikzpicture}[remember picture,overlay]
\node[anchor=south,yshift=10pt] at (current page.south) {\fbox{\parbox{\dimexpr\textwidth-\fboxsep-\fboxrule\relax}{\copyrighttext}}};
\end{tikzpicture}%
}

\usepackage{authblk}
\title{\LARGE \bf
Assessment of Cyberattack Detection-Isolation Algorithm\\ for CAV Platoons Using SUMO
}

\author[1]{Sanchita Ghosh}
\author[1]{Tanushree Roy}
\affil[1]{Department of  Mechanical Engineering, Texas Tech University, Lubbock, TX 79409, US. Emails:~{\tt\small sancghos@ttu.edu, tanushree.roy@ttu.edu}.}

\begin{document}

\copyrightnotice

\maketitle
\thispagestyle{empty}
\pagestyle{empty}


\begin{abstract}

 A Connected Autonomous Vehicle (CAV) platoon in an evolving real-world driving environment relies strongly on accurate vehicle-to-vehicle (V2V)  and vehicle-to-infrastructure (V2I) communication for its safe and efficient operation. However,  a cyberattack on this communication network can corrupt the appropriate control actions, tamper with system measurement, and drive the platoon to unsafe or undesired conditions. As a first step toward practicable resilience against such V2V-V2I attacks, in this paper, we implemented  a unified V2V-V2I cyberattack detection scheme and a V2I isolation scheme for a CAV platoon under changing driving conditions in Simulation of Urban MObility (SUMO). The implemented algorithm utilizes vehicle-specific residual generators that are designed based on analytical disturbance-to-state stability, robustness, and sensitivity performance constraints. Our case studies include two driving scenarios where highway driving is simulated using the Next-Generation Simulation (NGSIM) data and urban driving follows the benchmark EPA Urban Dynamometer Driving Schedule (UDDS). The results validate the applicability of the algorithm to ensure CAV cybersecurity and demonstrate the promising potential for practical test-bed implementation in the future.

\end{abstract}
\section{INTRODUCTION}
Intelligent Transportation System (ITS) equipped with adaptive traffic controls and CAVs is gradually becoming an essential part of modern transportation by providing improved efficiency, safety, and energy sustainability \cite{dey2015review}. ITS relies heavily on V2V, V2I, and vehicle-to-everything (V2X) (i.e. both V2V and V2I) communications to ensure reliable and energy-efficient operations of CAVs. Furthermore, vehicular ad-hoc network (VANET) is one of the most explored technologies that can support these communications along with several required facilities and protocols \cite{vanet,wyk2020}. While such communications are essential to incorporate safety-critical information for effective platoon control, they also increase the potential risks of threats \cite{buinevich2019forecasting, roy2023redundancies}. Although detecting-isolating cyberattacks is crucial in general, it is especially challenging for multi-modal systems such as CAVs \cite{ghosh2024koopman, sanchita_CCTA2023security}.

Over the last decade, cybersecurity of CAVs and VANET has received a significant amount of attention in the ITS research sector \cite{roy2020secure, hussain2020trust}. For instance, \cite{kalinin2017network, sdn} focused on information-based safety analysis while investigating secured network architectures, \cite{dataencrypt, trustdata} investigated reliable data transmission, and several other works investigated both data-driven and model-based cyberattack detection and resilient control for CAVs to minimize the risk severity posed as and when the informatics-based security measures are breached [vide references within \cite{survey, sedar2023comprehensive}]. Apart from analyzing such detection and corrective measures for CAV security, researchers also investigated attack model synthesis to assess the potential cyberthreats and their impacts on CAVs \cite{wang2020modeling}. In \cite{wang2024analytical, yang2023risk, dong2020impact}, authors devised and analyzed V2V cyberattack policies and demonstrated their effect on traffic flow and energy efficiency for CAVs. Moreover, the authors modeled and illustrated the impact of several V2V cyberattack strategies using realistic microscopic traffic simulator SUMO and network simulator OMNET++ in \cite{singh2018impact}. Similarly, authors conducted rigorous experiments utilizing SUMO-based traffic simulators to investigate and illustrate the impact of V2I attacks on CAVs in \cite{ramsamooj2024genvram, ekedebe2015simulation}. 

Nevertheless, the security concerns of CAVs with multi-modal controllers have remained underexplored. While a multi-mode control strategy in CAVs is essential to ensure optimal operation such as string stability, energy efficiency, vehicle safety, etc. under changing driving environments \cite{ploeg2011design,song2018multi,zhai2018switched},  this switching between various modes of operations significantly enhances the risks of compromised switching attacks on CAV operation \cite{sanchita_CCTA2023security}. To address this research gap, in our previous work \cite{ghosh2023cyberattack}, we proposed a V2X detection and V2I isolation scheme for a multi-modal CAV platoon. In particular, \cite{ghosh2023cyberattack} focuses on designing the measurement-based detector-isolator algorithms, derivation of the analytical guarantees to ensure benchmark performance, and demonstrating the performance of the proposed algorithms using simplistic car-following CAV platoon models. Consequently, validation of our proposed algorithm using a traffic microsimulator such as SUMO is the essential first step towards its successful practical implementation in the future. 
Thus, in this paper, our contributions are the following.
\begin{enumerate}
    \item We assess the impact of different V2V and V2I cyberattack policies on realistic CAV driving scenarios by implementing a high-fidelity experimental setup of a ring road network in SUMO.
    \item We design our detection-isolation algorithm \cite{ghosh2023cyberattack} for this CAV platoon using a rate-limited measurement sampling frequency of 20 Hz, congruent to a realistic Global Positioning System (GPS) update.
\end{enumerate}

The rest of the paper is as follows. Section~\ref{prob} describes the problem statement for the CAV model and the V2V-V2I cyberattack policies. Section~\ref{di_scheme} introduces the detection-isolation algorithm. Section~\ref{SUMO_sec} presents the implementation in SUMO and evaluates the performance of the algorithm for three case studies. We conclude the paper in Section~\ref{conclu}. 

\section{PROBLEM STATEMENT} \label{prob}
\subsection{Framework and CAV model}
Let us consider a car-following vehicle platoon with a string of one leader and $n$ follower vehicles under cooperative adaptive cruise control (CACC) with V2V and V2I communication. In this framework, the human-driven or semi-autonomous leader vehicle drives in response to the surrounding traffic environment, and the autonomous follower vehicles follow their preceding vehicle \cite{yao2020managing}. Furthermore, each follower vehicle adopts a controller based on the car-following model that attempts to reach a desired headway using the V2V information exchange \cite{ploeg2011design}. However, as the desired headway changes with the evolving driving environment, a multi-modal controller corresponding to a different driving mode is required to ensure the platoon's string stability and energy-efficient traffic flow \cite{song2018multi}. 
In particular, a supervisory controller monitors the trajectory of the leader vehicle to select the appropriate operational mode or driving scenario and then activates the corresponding vehicular controller gain through V2I communication \cite{turri2016cooperative}.
Now, to model this CAV platoon, let us define an augmented state vector $\chi_i = \begin{bmatrix}
    h_i & v_i & a_i & u_i
\end{bmatrix}^T$, where $h_i, v_i, a_i,$ and $u_i$ are respectively headway, velocity, acceleration, and control input of the $i^{th}$ vehicle. Here, $i\in\{0, \cdots,n\}$, where $n+1$ is the total number of vehicles in the platoon and $i =0$ indicates the leader vehicle. Then, we 
define the closed-loop platoon dynamics $\forall i\in\{1, \hdots,n\}$ as follows \cite{ploeg2011design}:
\begin{equation}\label{single_aug}
    \dot{\chi}_i = A_\alpha \chi_i + D_{\phi \alpha} \chi_{i-1} - S_{\alpha} s_i + \omega_i, \hspace{3mm} y_i = C \chi_i. 
\end{equation}
 Here, $s_i$ is the standstill distance, $\omega_i$ is the uncertainty, and $y_i = h_i \in \mathbb{R}$ is the headway of the $i^{th}$ vehicle. Additionally, $\alpha : [0,\infty) \rightarrow \mathcal{M} = [1, 2, \cdots, m]$ is the piecewise constant switching signal sent from the supervisory controller through the V2I network and indicates the operational mode of the platoon among the $m$ number of admissible driving modes for the platoon. The function $\phi : [0,\infty) \rightarrow \{ 0, 1 \}$ represents V2V communication where the values $0$ and $1$ indicate the ceased or active communication respectively and $\phi_i \chi_{i-1}$ denotes the data received over V2V for the $i^{th}$ vehicle.
 $A_\alpha,D_{\phi \alpha}$,  $C$, and $S_{\alpha}$ are system matrices defined in \eqref{matP}  based on the standard car-following model. Here, $k_{1,\alpha}, k_{2,\alpha}, k_{3,\alpha} $, and $ k_{u,\alpha}$ are the suitable vehicular controller gains and $T_\alpha$ is the desired headway at platoon operational mode $\alpha$  [refer to \cite{ghosh2023cyberattack} for details on these matrix constructions]. 
%
\begin{align}\label{matP}
   & A_\alpha = \begin{bmatrix}
        0 & -1 & 0 & 0\\
        0 & 0 & 1 & 0 \\
        0 & 0 & -\frac{1}{\Sigma}& \frac{1}{\Sigma} \\
        k_{1,\alpha} & k_{2,\alpha} & k_{3,\alpha}& k_{u,\alpha}
    \end{bmatrix}; \quad C = \begin{bmatrix}
            1 \\ 0 \\  0 \\ 0
        \end{bmatrix}^T;  \\
    & D_{\phi \alpha} = \begin{bmatrix}
        0 & 1 & 0 & 0\\
        0 & 0 & 0 & 0 \\
        0 & 0 & 0 & 0\\
        0 & \overline{k}_{2,\alpha} & \overline{k}_{3,\alpha}& \frac{\phi_i}{T_\alpha} 
    \end{bmatrix};  \quad S_{\alpha} = \begin{bmatrix}
            0 \\ 0 \\ 0 \\ k_{1,\alpha}
        \end{bmatrix}.
\end{align}
\subsection{Cyberattack policy} \label{sec:policy}
\begin{figure}[h!]
    \centering
    \includegraphics[trim = 0mm 0mm 0mm 0mm, clip,  width=0.6\linewidth]{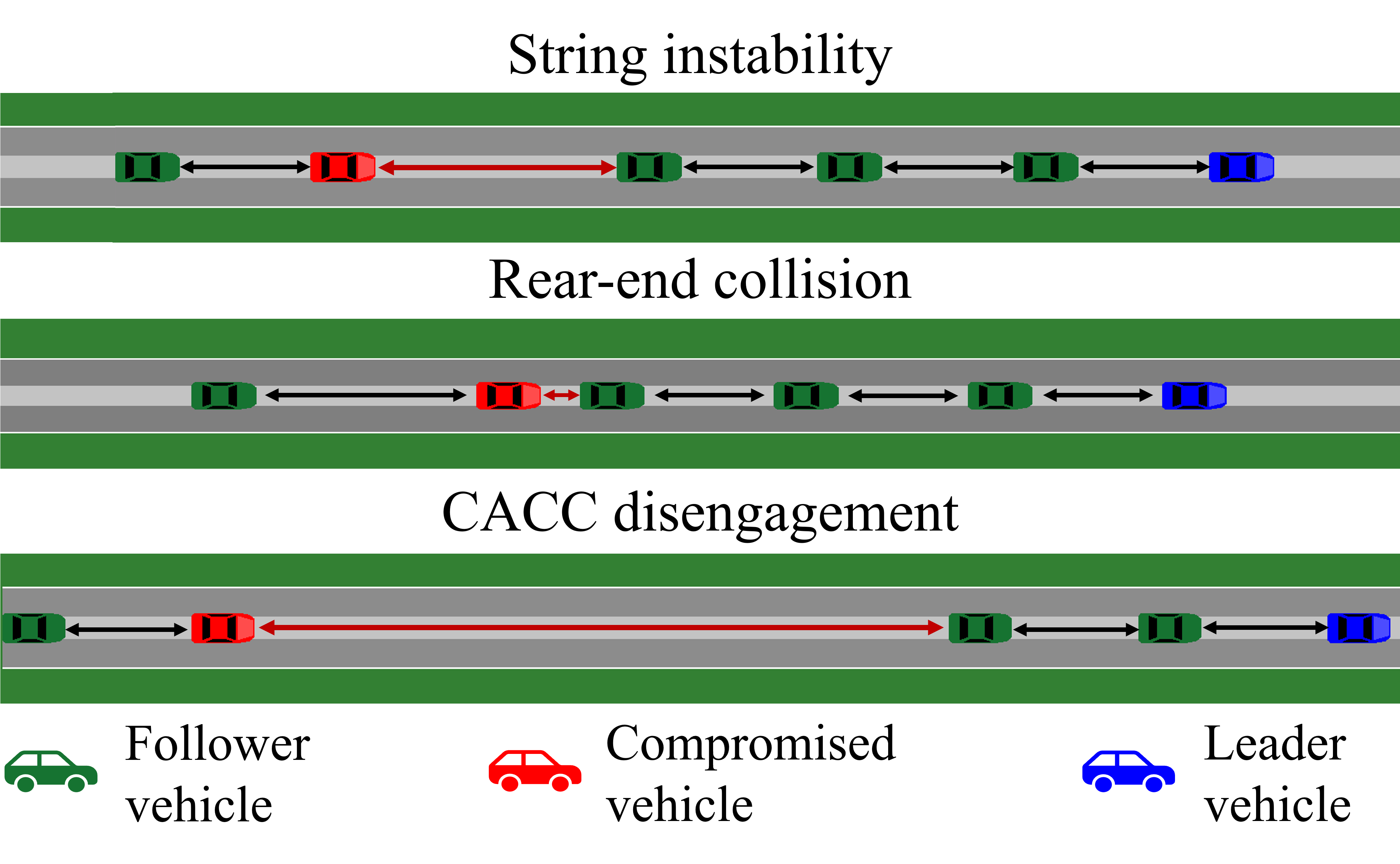}   
    \caption{Typical impacts of V2V and V2I attacks on CAV platoon.}
    \label{fig:comb}
\end{figure}

In this framework, the adversary can attack single or multiple follower vehicles through the V2I and/or through the V2V communication channel to drive the platoon to unsafe or undesired traffic conditions. For example, an adversary may induce a brisk velocity perturbation to achieve string instability in the platoon to exacerbate travel time and fuel consumption \cite{ekedebe2015simulation}. This scenario is shown in the top plot of \mbox{Fig. \ref{fig:comb}}. During unsafe conditions, compromised vehicles may run with a very small headway distance (too close to the preceding vehicle), and this can lead to a high probability of collisions during stop-go traffic conditions, as shown in the middle plot of \mbox{Fig. \ref{fig:comb}}. 
Alternatively, an unstable compromised vehicle may run with a larger headway and a too low velocity compared to the preceding vehicle. The vehicle may eventually trail off if the headway distance becomes too large, leading to a disengaged CACC \cite{gunter2020commercially}.  The bottom plot of \mbox{Fig. \ref{fig:comb}} exhibits this scenario. 

Specifically, we consider that the adversary attempts to tailor a V2I attack under a changing driving environment scenario to mimic a compromised switching attack scenario, and such an attack exhibits a high likelihood of leading the system to an unsafe and unstable operating region \cite{sanchita_CCTA2023security}. 
In addition, V2I attack scenarios may compromise the communications between all vehicles and infrastructure components. Consequently, the adversary may access and manipulate multiple CAVs simultaneously to initiate large-scale traffic disruptions and create complex adversarial traffic patterns through infrastructure-level tampering such as creating catch-me-if-you-can scenarios, targeted congestion, etc. \cite{reilly2016creating,ghena2014green, roy2023redundancies}. Therefore, in this work, we focus on isolating the V2I attack scenarios.
Furthermore, the adversary can opt for several active network attacks, e.g., denial-of-service (DoS),  distributed DoS, false-data-injection (FDI), blackhole, and replay attacks through both V2V and V2I communication to achieve the attack goals shown in \mbox{Fig. \ref{fig:comb}}. 
Under these potential attacks, the compromised platoon dynamics is defined:
\begin{align} \label{zeta}
     \dot{\chi}_i = & A_\alpha \chi_i + D_{\phi \alpha} \chi_{i-1} - S_{\alpha} s_i + E_{\phi i} f_{\phi i} + E_{\alpha i} f_{\alpha i}  \nonumber \\ & + E_{\phi \alpha i} f_{\phi \alpha i} + \omega_i, \quad \chi_i(0)=\chi_{i0}, \quad t\geqslant 0.
\end{align}
 Here, $ E_{\phi i}, E_{\alpha i}, E_{\phi \alpha i}$ are matrices and $f_{\phi i}, f_{\alpha i}, f_{\phi \alpha i}$ are time functions representing only V2V, only V2I, and V2X (simultaneous V2V-V2I) attacks, respectively.


\section{DETECTION- ISOLATION ALGORITHM} \label{di_scheme}
We propose a two-phase algorithm to detect the V2X attacks and subsequently, isolate the 
V2I attacks while identifying the compromised vehicles. 
\begin{enumerate}
    \item \textbf{V2X detection scheme (DS):} First, the V2X DS monitors individual vehicle trajectories and utilizes V2X DS residual to capture any deviation from the desired behaviors. In particular, V2X DS only uses the headway $h_i$, while $E_{\phi i} f_{\phi i}, \, E_{\alpha i} f_{\alpha i}$, and $E_{\phi \alpha i} f_{\phi \alpha i}$ remain unknown. A V2X attack is detected when the V2X DS residual crosses a predefined threshold $J_{DS}$. 
    \item \textbf{V2I isolation scheme (IS):} Upon detection of a V2X cyberattack,  the second phase of the algorithm, the V2I IS is activated for that vehicle. The V2I IS acquires the V2V communicated data $\phi_i \chi_{i-1}$ and investigates the vehicle's performance under $\phi_i \chi_{i-1}$ to ascertain if the performance deterioration is due to a V2I attack. Since $\phi_i \chi_{i-1}$ captures the V2V corruption (if present), $E_{\phi i} f_{\phi i}$ remains known to the V2I IS. Similarly, it utilizes V2I IS residual such that the residual crosses a threshold $J_{IS}$ under the presence of a V2I attack.
\end{enumerate} 
Furthermore, both V2X DS and V2I IS act on individual vehicles which enables the algorithm to identify the compromised vehicles and thus localize the source of attack. 
Hence, we consider banks of $n$ numbers of switched-mode detectors and isolators corresponding to the $n$ follower vehicles of the platoon. All these $2n$ detectors-isolators have $m$ operational modes corresponding to the $m$ admissible platoon modes and run with the same mode $\alpha$ at a time. The algorithm receives the mode information from the supervisory controller. 
Mathematically,  we define the detector-isolator dynamics  as:
\begin{align}
    \text{V2X DS:} \quad & \dot{\hat{\chi}}_i =  A_\alpha \hat{\chi}_i + D_{\phi \alpha} \hat{\chi}_{i-1} - S_{\alpha} s_i +L_\alpha (y_i-\hat{h}_i),\nonumber \\
    & \hat{h}_i = C \hat{\chi}_i; \quad \forall i \in \{ 1, 2, \cdots,n\}, \quad t \geqslant 0. \label{chihat} \\
    \text{V2I IS:} \quad & \dot{\hat{\psi}}_i  =  A_\alpha \hat{\psi}_i  + D_{\phi \alpha} \hat{\psi}_{i-1} - S_{\alpha} s_i +  E_{\phi i} f_{\phi i} \nonumber \\ & +M_\alpha (y_i-\hat{H}_i), \quad \hat{H}_i  = C \hat{\psi}_i, \quad t \geqslant t_0, \label{chihat2}
 \end{align} where $t_0$ is the time of V2X attack detection and $\hat{\psi}_i(t_0) = \hat{\chi}_i (t_0)$.
Here $\hat{\chi}_i, \hat{\psi}_i \in \mathbb{R}^{4}$ and $\hat{h}_i, \hat{H}_i \in \mathbb{R}$ are respectively detector-isolator states and outputs (estimated headway).   $L_\alpha, M_\alpha \in \mathbb{R}^{4 \times q}$ are the  detector-isolator gain matrices to be designed.  Theorem 1 in \cite{ghosh2023cyberattack} provides the analytical guarantees on $L_\alpha, M_\alpha$ design constraints to ensure standard diagnostic performance criteria such as exponential stability, disturbance-to-state stability, robustness against uncertainties, and sensitivity towards cyberattacks. Next, we monitor the deviation in vehicle headway to first detect V2X attack and then isolate V2I attack. Therefore, with detector-isolator estimated headways $\hat{h}_i$ and $\hat{H}_i$\normalsize,
we define:
\begin{align}
 &\overline{r}_i =  \left[ \lVert y_i - \hat{h}_i \rVert_2^2 - \lVert {y}_{i-1} - \hat{h}_{i-1} \rVert_2^2 \right]\label{dsRes0}\\
   & \text{V2X DS residuals:} \quad  r_{c, i} =   \max{\left(0, \, \overline{r}_i \right)}; \label{dsRes}\\
    &\text{V2I IS residual:} \quad \, \gamma_i \, =  \, y_i - \hat{H}_i. \label{isRes}
\end{align} We note here that, for leader vehicle ${\chi}_{0} - \hat{\chi}_{0} = 0$. Next, $r_{c, i} $ and $\gamma_i$ are compared with predefined thresholds $J_{DS}$ and $J_{IS}$, respectively, to make accurate detection-isolation decision such that if $r_{c, i} \geqslant J_{DS}$ then the $i$-th vehicle is the V2X inflicted vehicle and a flag is raised. On the other hand, if $\gamma_i  \geqslant J_{IS}$ then a \texttt{V2I Attack Flag} is generated, inferring the presence of V2I attack in the platoon. 
Nevertheless, appropriate threshold selection is crucial to ensure the reliable detection-isolation performance of the proposed algorithm. In particular, we intend to choose an optimum threshold value $J_{DS}$ to minimize mis-detection(-isolation) under V2X attack (V2I attack) while minimizing the false alarm rate due to uncertainty $\omega_i$ (only V2V attack) as well \cite{Ding}. 
Therefore, we evaluate and analyze the residuals $r_{c, i} $ and $\gamma_i$ in the absence of V2X and V2I attacks, respectively, to determine the suitable threshold values.
Now, we present the algorithm for detection and isolation in Algorithm~\ref{algo}. 

\begin{algorithm2e} \label{algo}
\caption{Generate \texttt{V2X Attack Flag}, \texttt{V2I Attack Flag}, and compromised \texttt{vehicle ID}}\label{alg:desicion}
\KwIn{Time instant $t$, measurement $y_i$, number of vehicles $n$, initial conditions $\hat{\chi}_i(0),\hat{\psi}_i(0)$, operation mode $\alpha$, detector \& isolator parameter set $P_\alpha=[A_\alpha, D_{\phi \alpha}, S_{\alpha},  L_\alpha,M_\alpha]$, thresholds $J_{DS},J_{IS}$,V2V  data $\phi_i \chi_{i-1}$}
\KwOut{V2X Attack Flag with Vehicle ID.}
\SetKwFunction{FD}{Detector}
\SetKwFunction{FI}{Isolator}

    \For{$t\geqslant0$}{
  \For{$i=1:n$}{
    $r_{c, i} \leftarrow$ \texttt{Detector($i,y_i,\hat{\chi}_i(0),P_\alpha$)}\\
    \If{$r_{c, i} \geqslant J_{DS}$}{
    \texttt{vehicle ID}=i;\\
    \KwRet \texttt{ V2X Attack Flag,vehicle ID}\\
    $\gamma_{i} \leftarrow$ \texttt{Isolator($i,y_i,\hat{\psi}_i(0),P_\alpha,\phi_i \chi_{i-1}$)}\\
    \If{$\gamma_i  \geqslant J_{IS}$}{
    \KwRet \texttt{ V2I Attack Flag} \;}
    }}
    }

    \SetKwProg{Fn}{Function}{:}{\KwRet}
    \Fn{\FD{$i,y_i,\hat{\chi}_i(0),P_\alpha$}}{
        Evaluate $\hat{\chi}_i$ using \eqref{chihat} and then find $\hat{h}_i$.\\
    Compute residual $r_{c, i}(t)$ using $h_i(t)$ \eqref{dsRes0}-\eqref{dsRes}.\\
        \KwRet $r_{c, i}(t)$ \;
        }

    \SetKwProg{Fn}{Function}{:}{\KwRet}
    \Fn{\FI{$i,y_i,\hat{\psi}_i(0),P_\alpha,\phi_i \chi_{i-1}$}}{
        Evaluate $\hat{\psi}_i$ using \eqref{chihat2} and  find $\hat{H}_i(t)$.\\
        Compute residual $\gamma_{i}(t)$ using $h_i(t)$, \eqref{isRes}.\;
        \KwRet $\gamma_{i}(t)$ \;
        }
\end{algorithm2e}

%
\section{SUMO VALIDATION}\label{SUMO_sec}
The focus of our work lies in validating our detection-isolation Algorithm~\ref{algo} using the SUMO platform. Specifically, SUMO is an open-source microscopic traffic simulator that supports individual vehicle control and behavior simulation on user-defined road networks and traffic demand \cite{SUMO2018}. Thus, we implement our road topology, CAV platoon operations, on-board controllers, and two driving environments, namely highway and urban driving, in this platform. Furthermore, in this SUMO-based high-fidelity CAV simulator, we demonstrate the impact of various V2V and V2I attack policies (as described in Section~\ref{sec:policy}).  
\begin{figure}[h!]
    \centering
    \includegraphics[trim = 0mm 0mm 0mm 0mm, clip,  width=0.5\linewidth]{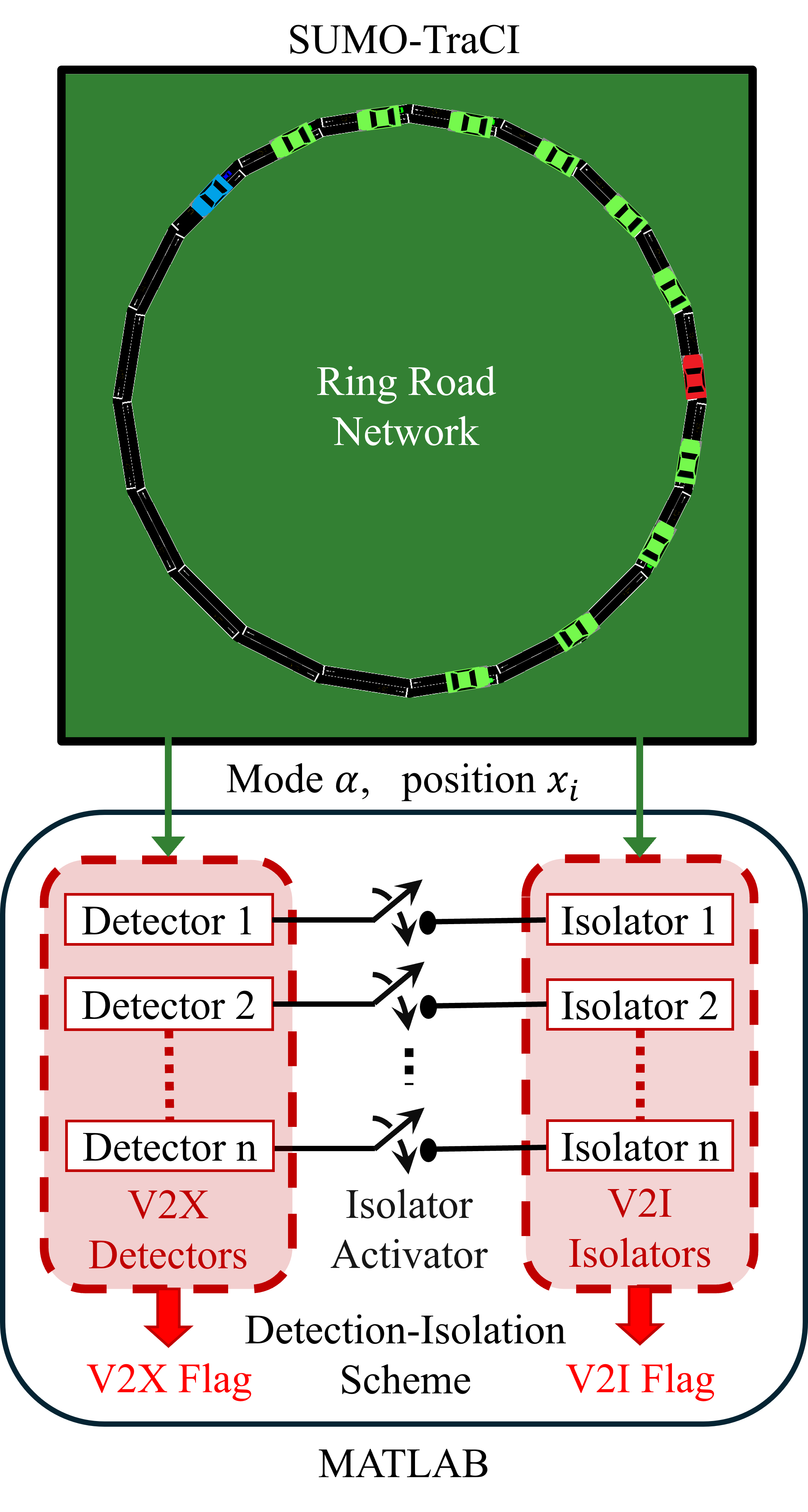}   
    \caption{Block diagram showing the communication between the SUMO-TraCI platform (representing the road topology, driving environment, CAV platoon and on-board controllers) and Matlab (representing the supervisory controller and detector-isolator system) for the simulation case studies. }
    \label{fig:simsetup}
\end{figure}

We utilize SUMO's Python API, the TraCI (Traffic Control Interface), which gives users access to the traffic simulation environment. We use TraCI to first retrieve the individual vehicle's states, i.\,e., position, velocity, and acceleration. Then, we calculate the desired velocity for the CAVs with CACC input using \eqref{zeta} and subsequently, use TraCI to provide the controller commands.  TraCI essentially behaves as the V2V and V2I communication and thus enables us to inject different attack policies into the platoon. 
Now, to detect V2X attacks and isolate the presence of V2I attacks, we utilize the headway and velocity measurements of the vehicles from the TraCI. In particular, we send the measurement  and the mode information to the Matlab-based detection-isolation scheme, following Algorithm~\ref{alg:desicion}. Fig.~\ref{fig:simsetup} shows our simulation setup connecting the SUMO-TraCI configuration that represents the road topology, driving environment, CAV platoon, and on-board controllers, and Matlab implementation that represents the supervisory controller and detector-isolator system. We present further details of our simulation setup below.

\vspace{3mm}
 \noindent 
\textbf{Road Network:}  Here, we consider a circular road network to conduct the Ring Road experiment \cite{sugiyama2008traffic}. The ring consists of 20 links and a total length of 600m. 

\noindent
\textbf{Traffic Demand:}
We simulate a platoon of 12 vehicles with one leader vehicle and 11 follower CAVs. The leader vehicle runs with a given velocity trajectory and the follower CAVs runs with the CACC input.

\noindent
\textbf{Simulation Timestep:}
The car-following models have been implemented for a simulation timestep of $50 ms$ in SUMO to evaluate the proposed algorithm's performance under a realistic measurement sampling rate of 20 Hz through the GPS (global positioning system) \cite{shu2022high}. 

\noindent
\textbf{Driving Environment:} For the simulation case studies, we consider two driving environments: highway and urban. Moreover, to further capture the complexities and uncertainties of real-world operations, we utilize the velocity trajectories from \href{https://ops.fhwa.dot.gov/trafficanalysistools/ngsim.htm}{NGSIM} and  \href{https://www.epa.gov/emission-standards-reference-guide/epa-urban-dynamometer-driving-schedule-udds}{UDDS} traffic datasets to simulate highway and urban driving environment respectively.

\noindent
\textbf{Controller Parameters:} We adopt the control scheme proposed in \cite{ploeg2011design} to obtain the vehicular controller gains. The vehicle controllers have gain values $k_1 = \begin{bmatrix}
    0.2 & 1.5 & 2
\end{bmatrix}$ and the desired time-headway $0.5s$ for  nominal highway driving. On the other hand, the vehicle controllers have gain values $k_2 = \begin{bmatrix}
    0.2 & 2 & 0
\end{bmatrix}$ and the desired time-headway $1.3s$ for nominal urban driving. Notably, the desired headway for urban driving is larger than that of highway driving due to the frequent stop-and-go behavior of urban traffic.

\noindent
\textbf{Detector Parameters:} for highway driving environment,  we choose  the detector gains, $L_1 $ and isolator gains $M_1$ as: \mbox{ $L_1 = M_1 =  \begin{bmatrix}
    0.2 & 0.2 & 0.4 & 0.8 \\
    0.2 & -0.2 & -0.2 & 0.8
\end{bmatrix}$.}
Similarly, we choose to be $L_2 = M_2 =  \begin{bmatrix}
    0.2 & 0.2 & 1.2 & -1.2 \\
    0.2 & 1.2 & -1.8 & -1.2
\end{bmatrix}$, for the urban driving environment. Additionally, we choose separate thresholds for these two different modes. For highway environment, we choose $J_{DS} = 4$ and $J_{IS} = 6$ and for urban environment, we choose $J_{DS} = 1$ and $J_{IS} = 3$.
%
Now, we present three case studies to demonstrate the impact of the attacks and evaluate the performance of our algorithm.
\newpage
\subsection{Case Study I: FDI attack through V2V network}
\begin{figure}[h!]
    \centering
    \includegraphics[trim = 0mm 0mm 0mm 0mm, clip,  width=0.6\linewidth]{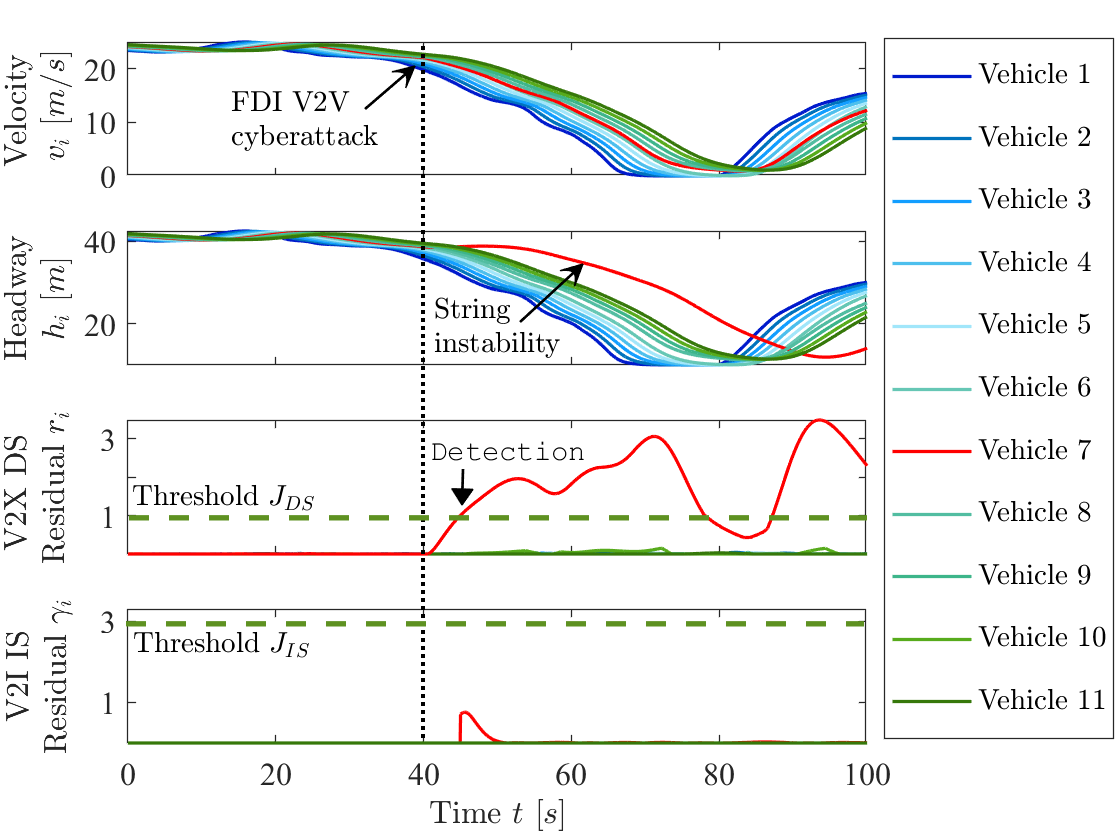}   
    \caption{Under V2V attack, the plot shows (top)  the velocity, (second) the position, (third) the V2X DS residual, and (last) the V2I IS residual. }
    \label{fig:v2v}
\end{figure}

We consider that the platoon is running in an urban traffic in this scenario.  Then, we introduce an FDI attack from the 40s through the V2V network for the  $8^{th}$ vehicle such that the vehicle tracks wrong desired acceleration and ultimately leads to string instability.  This is evident from the top two plots of Fig~\ref{fig:v2v} showing the headway and velocity of the vehicles in the platoon. Furthermore, the V2X residual of the vehicle 8  $r_{c, 8}$ crosses the threshold $J_{DS}$ and accurately generates the attack flag within 8s of attack as shown in the $3^{rd}$ plot of Fig~\ref{fig:v2v}. After the detection, the V2I IS is activated. However, since the platoon is under a V2V attack,  the V2I IS residual $|\gamma_8|$ does not cross the threshold $J_{IS}$ in the absence of a V2I attack. This is shown in the bottom plot of Fig~\ref{fig:v2v}.

\subsection{Case Study II: FDI attack through V2I network} 

\begin{figure}[h!]
    \centering
    \includegraphics[trim = 0mm 0mm 0mm 0mm, clip,  width=0.6\linewidth]{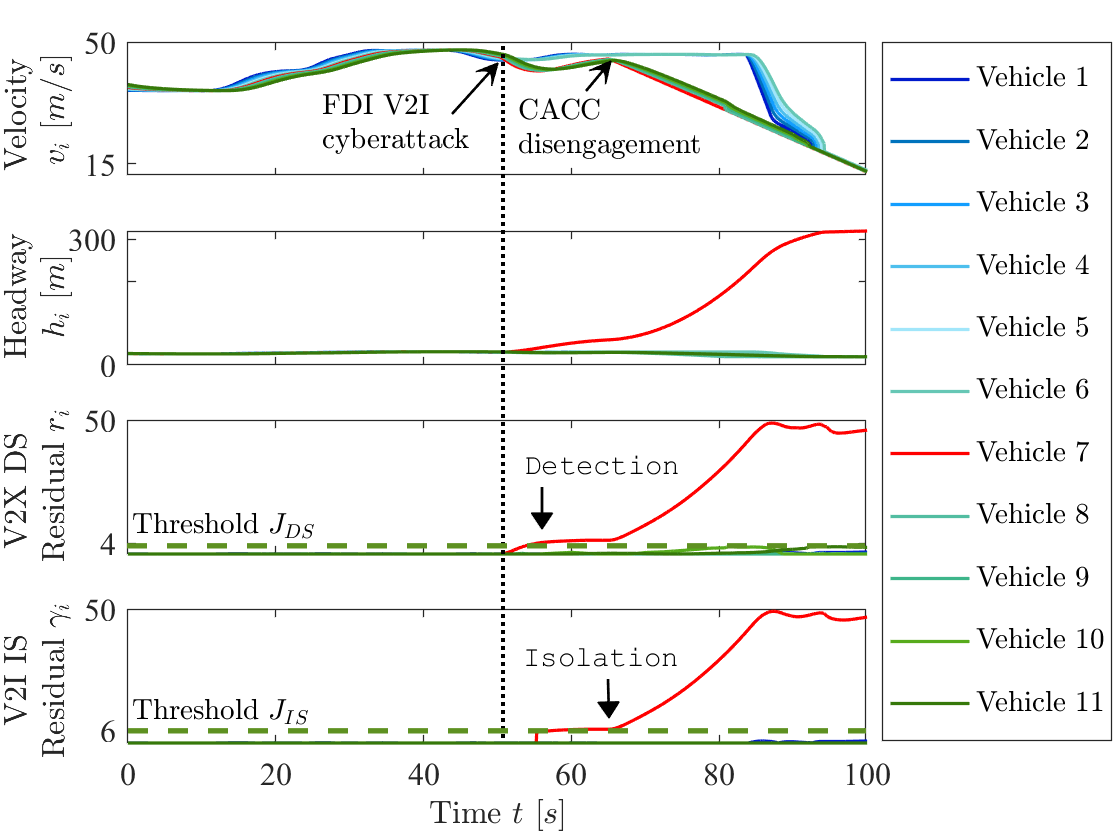}   
    \caption{Under V2I attack, the plot shows (top)  the velocity, (second) the position, (third) the V2X DS residual, and (last) the V2I IS residual. }
    \label{fig:diseng}
\end{figure}
In the scenario, we simulate the platoon running on a highway under a FDI V2I attack in the SUMO environment. The attack is injected to the $8^{th}$ vehicular controller from 20s such that the controller mode signal $\alpha$ is altered for this vehicle. 
Thus, the $8^{th}$ vehicle effectively runs with the  urban driving environment's controller gains and the desired headway in the highway driving environment. Consequently,  the vehicle starts to lag behind the $7^{th}$ vehicle further with time. This ultimately leads to CACC disengagement because of the vast distance between the $7^{th}$ and $8^{th}$ vehicles.
The rest of the CAVs that are following the $8^{th}$ vehicle are also disengaged. However, due to circular nature of the ring road network of our simulation, eventually all the vehicles start to halt as shown in the top two plots of \mbox{Fig. \ref{fig:diseng}}  showing the headway and velocity of the vehicles in the platoon, which is also expected from previous experimental studies \cite{gunter2020commercially}. Now, the V2X residual of the vehicle 8  $r_{c, 8}$ crosses the threshold $J_{DS}$ within 6s of attack occurrence to accurately detect the attack as shown in the third plot of \mbox{Fig. \ref{fig:diseng}}.
After a V2X detection, the $8^{th}$ V2I isolator is activated. Furthermore, as the platoon is under V2I attack, the V2I IS residual $|\gamma_8|$ crosses the threshold  $J_{IS}$ and isolates the presence of the V2I attack within 14s of activation (shown in last plot \mbox{Fig. \ref{fig:diseng}}).

 \subsection{Case Study III: DoS attack through V2I network} 

\begin{figure}[h!]
    \centering
    \includegraphics[trim = 0mm 0mm 0mm 0mm, clip,  width=0.6\linewidth]{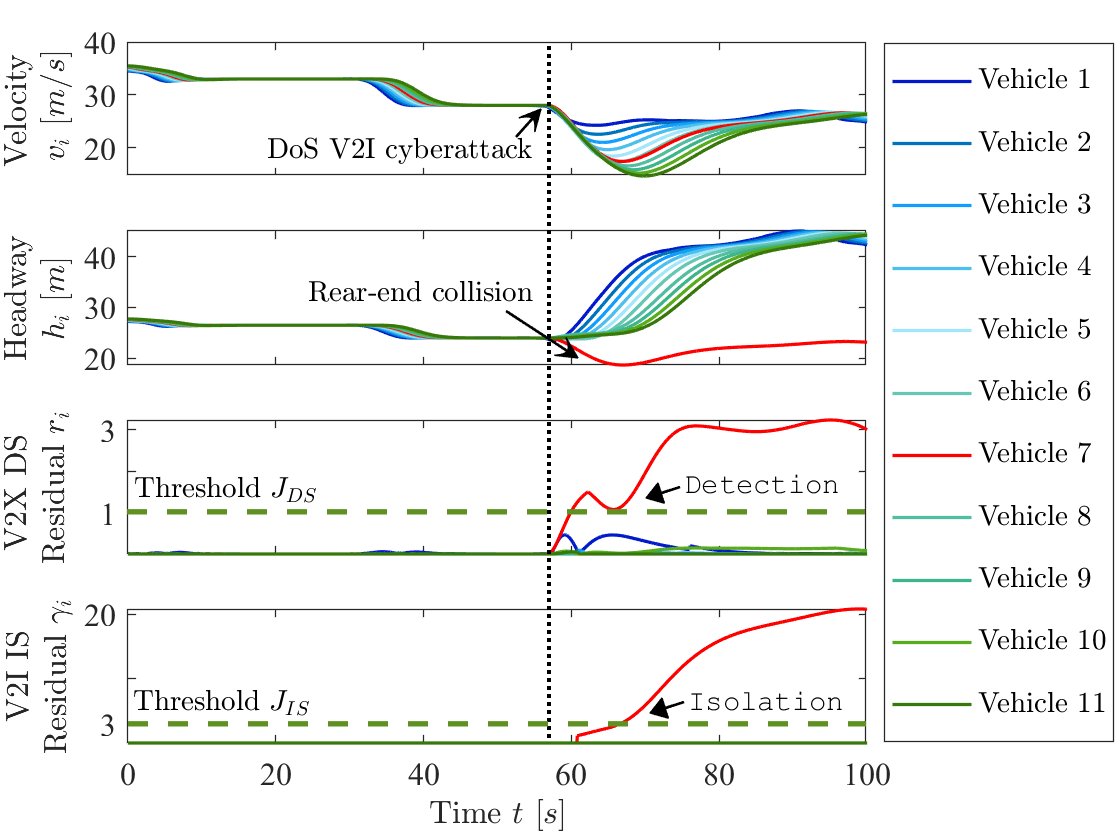}   
    \caption{Under V2I attack, the plot shows (top)  the velocity, (second) the position, (third) the V2X DS residual, and (last) the V2I IS residual. }
    \label{fig:v2iud}
\end{figure}
Here, we consider that the vehicle platoon is entering an urban traffic from highway environment at 57s. Then, the adversary jams the V2I network for the $8^{th}$ vehicle to introduce a DoS attack. Consequently, the $8^{th}$ vehicle fails to receive the updated control command and continues to run with highway driving control settings in an urban driving environment. This causes the vehicle to drive too closely to the $7^{th}$ vehicle, i.\,e.  the $8^{th}$ vehicle runs with a very small headway. This phenomenon shows a high possibility of the rear-end collision. This is evident from the top two plots of \mbox{Fig. \ref{fig:v2iud}}  showing the position and velocity of the vehicles in the platoon. Now, the V2X residual of the vehicle 8  $r_{c, 8}$ crosses the threshold $J_{DS}$ within 5s of attack occurrence to accurately detect the attack as shown in the third plot of \mbox{Fig. \ref{fig:v2iud}}. After a V2X detection, the $8^{th}$ V2I isolator is activated. Furthermore, as the platoon is under V2I attack, the V2I IS residual $|\gamma_8|$ crosses the threshold $J_{IS}$ confirming the presence of the V2I attack within 8s of activation. This is shown in the last plot in \mbox{Fig. \ref{fig:v2iud}}.

\section{CONCLUSIONS}\label{conclu}
In this work, we validated a two-phase  V2V and V2I cyberattack detection-isolation schemes for CAV platoon by implementing in SUMO-TraCI and Matlab combined platform. The cyberattacks were implemented through the TraCI communication and their impact showed common CAV disruptions such as CACC disengagement or rear-end collision. We designed our set of model-based detectors and isolators for each vehicle to ensure stability with optimal robustness and sensitivity guarantees. Furthermore, we demonstrated the effectiveness of our algorithm using simulation case studies for three attack scenarios: only V2V, only V2I, and simultaneous V2V-V2I. The CAV driving environments were simulated here using UDDS and NGSIM data along with common cyberattacks such as DDoS, FDI, and replay attacks. In all the considered attack policies and driving scenarios, our implemented algorithm successfully detected vehicles under both V2V and V2I cyberattacks and isolated the scenarios for V2I corruptions under a realistic measurement sampling frequency of 20 Hz. 



\newpage

\bibstyle{arxiv}
\bibliography{ref.bib}

\end{document}